\documentclass[summary]{ursi}
\usepackage{color}

\newcommand{\MH}{\color{black}}
\usepackage{subcaption}
\title{Radio Environment Map for Energy-Efficient User-Centric Cell-Free M-MIMO Network}

\author{Marcin Hoffmann\affref{ref1}, and Pawel Kryszkiewicz\affref{ref1}}

\affiliation{%
  \aff{ref1}{Poznan University of Technology, Poznan, Poland}
}

\begin{document}

\maketitle

\begin{abstract}
This paper proposes a Radio Environment Map (REM) for energy-efficient (EE) serving cluster formulation in a user-centric cell-free network. By incorporating the location of the user and the characteristics of the power amplifier, REM enables EE to be improved by up to 19\%.
\end{abstract}

\section{Introduction}

The concept of Radio Environment Maps (REMs) originates from the Dynamic Spectrum Access (DSA), where they were used to store information about spectrum utilization in certain locations~\cite{Kliks2017}. Later on, it was found that 5G and beyond networks can benefit from any type of location-dependent data to support their Radio Resource Management algorithms. This can be achieved by extending the network infrastructure with a dedicated REM module containing maps of, e.g., traffic distribution or network coverage~\cite{Tengkvist2017MultidimensionalRS}. Currently, the future 6G networks are expected to utilize more complicated signal processing techniques than the previous generations, aiming to improve users Quality of Service (QoS), while maintaining high Energy Efficiency (EE). An example of such a technology is User-Centric Cell-Free Massive Multiple-Input Multiple-Output (UCCF M-MIMO)~\cite{Demir_2021}. In UCCF M-MIMO multiple base stations (BSs), usually referred to as Access Points (AP), coordinated by the so-called Central Processing Unit (CPU), are jointly providing the service to the User Equipments (UEs). From this perspective, one of the crucial challenges is to determine how many APs should serve a given UE (known in the literature as serving cluster formulation~\cite{Demir_2021}), i.e., in some situations serving a UE by multiple APs might improve its received power or reduce interference, while in others can exploit APs computational capabilities and radio resources without any benefit to the users QoS. Moreover, proper serving cluster formulation depends on the UE location in the cell, and has an impact on the network EE, i.e., it potentially allows the use of available network hardware to transmit more data while using the same amount of power. Especially this might be the case while taking into account that the UCCF MMIMO Network can be built from APs equipped with Power Amplifiers (PAs) of different classes~\cite{Kryszkiewicz2023}. 

From this perspective, we propose to formulate serving clusters based on the spatial distribution of users, similarly to what we proposed for cell on/off switching~\cite {HOFFMANN2021232}. In detail, we propose REM, which provides the mapping between the UE location pattern
and EE related to the particular serving cluster configuration. {\MH This serves as a starting point for further REM-based optimization of the UCCF M-MIMO network, e.g., greedy selection of the serving cluster configuration associated with the highest EE, or more advanced Reinforcement Learning (RL) algorithms possible to be considered in the future.} Compared to the previous study, we incorporate the non-ideal characteristic of PA installed at APs into REM, {\MH which we demonstrate to have an impact on the EE, and configuration of serving clusters in UCCF M-MIMO.} Throughout the rest of the paper, we show the structure of REM, and example entries obtained from the 3D Ray-Tracer (3D-RT)-based system-level simulator of UCCF M-MIMO network~\cite{hoffmann2024}.

\section{REM for UCCF M-MIMO Network}

\begin{figure}[htbp]
  \centering
  \includegraphics[trim=5.5cm 2.25cm 5.5cm 4.5cm, clip=true,width=0.45\textwidth]{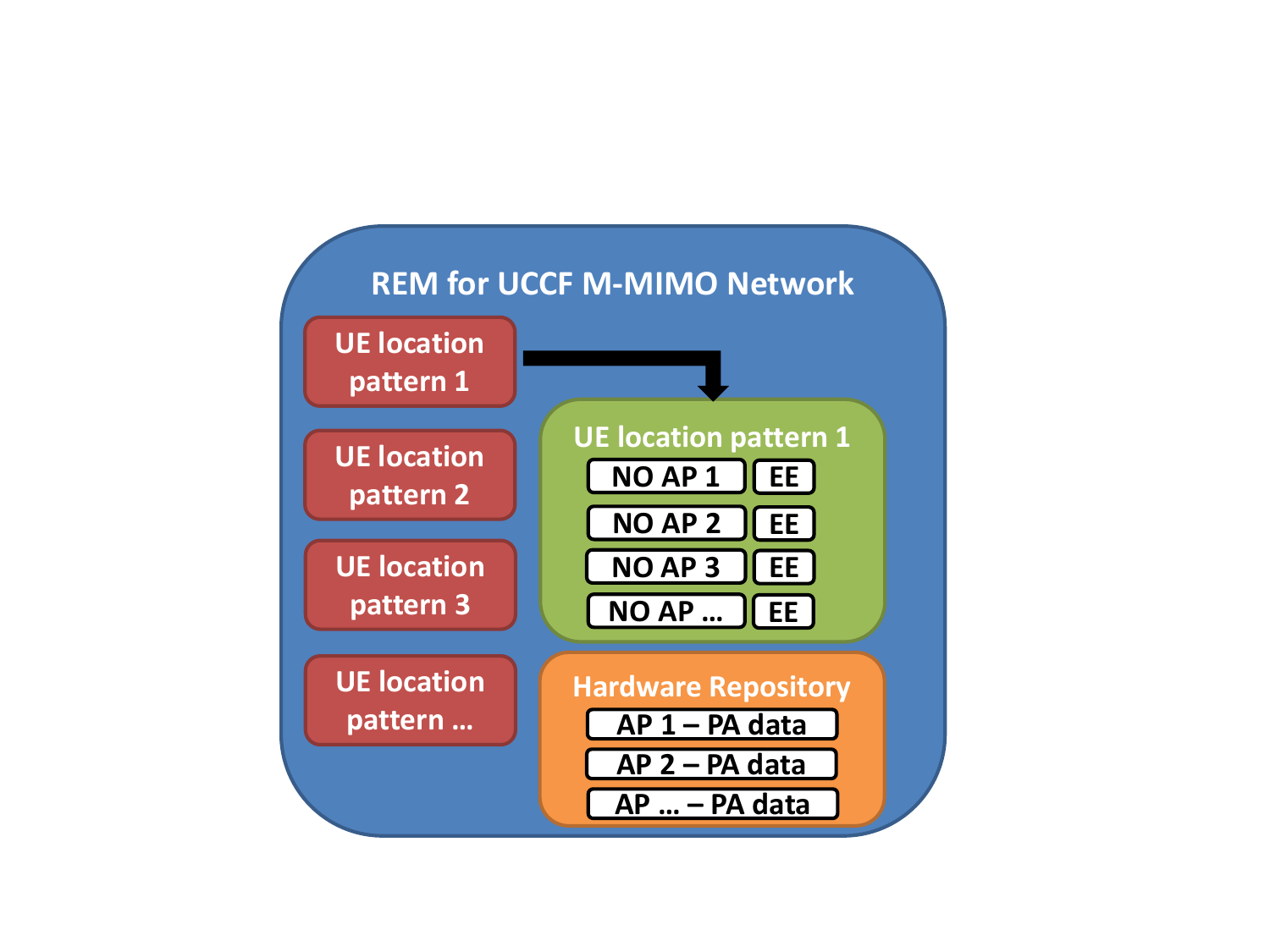}
  \caption{Proposed structure of REM for monitoring EE of UCCF M-MIMO network.}
  \label{fig:rem}
\end{figure}

\begin{figure}[htbp]
  \centering
  \includegraphics[trim=2.5cm 6.5cm 2.5cm 5cm, clip=true,width=0.45\textwidth]{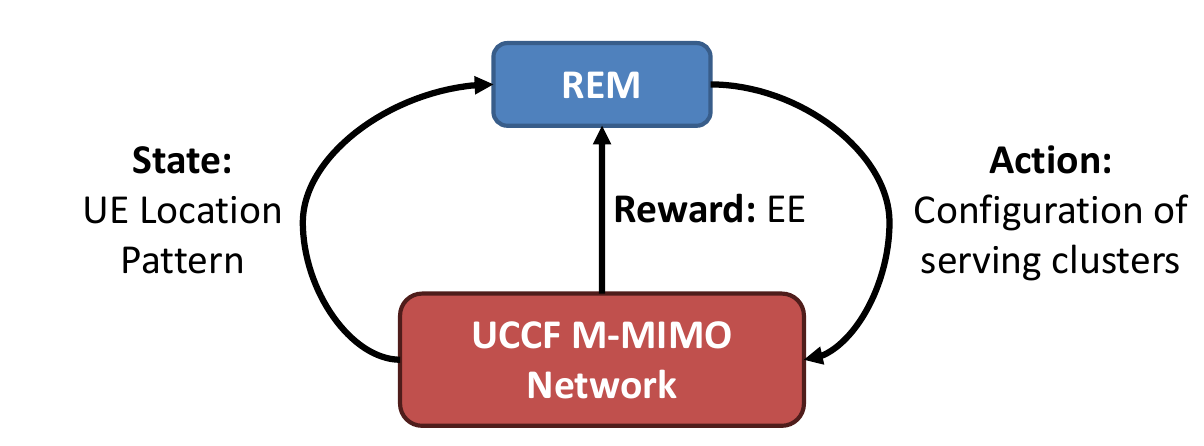}
  \caption{RL cycle for REM-based EE optimization of UCCF M-MIMO network.}
  \label{fig:rl}
\end{figure}

In this paper, we propose to extend the traditional network infrastructure with a REM dedicated to the optimization of UCCF M-MIMO network. We expect that REM is capable of collecting the UE location data (possibly using high-accuracy satellite positioning systems~\cite{HOFFMANN2021232}) along with network Performance Metrics (PMs) such as total data volume and energy consumption. In addition, we propose to incorporate in REM information about the hardware installed at APs, or more specifically, PAs, which are expected to consume the majority of power. {\MH We expect that these supplemental data can provide insights on the configuration of serving clusters, e.g., some PAs can generate more distortion, or consume more power, making related APs worse candidates for participating in large serving clusters compared to the APs equipped with energy-efficient hardware.} 
The data in REM is organized in entries labeled with the \emph{UE Location Pattern} being an array of $x,y$ coordinates of the UEs. {\MH For each \emph{UE Location Pattern}, REM provides an EE (defined as the ratio between total data volume and total energy consumption), associated with a given serving cluster configuration. Here, \emph{NO AP} represents the number of highest-power APs that should serve a UE. In addition, the REM is equipped with a \emph{Hardware Repository} (see Fig.~\ref{fig:rem}), which contains information about, e.g., the PA associated with each AP. This data is expected to be used for large-scale analysis, e.g., through data analysis, some dependencies can be observed between APs, PAs, and the serving cluster configuration providing the highest EE. Therefore, the proposed structure of REM allows one to analyze the best serving cluster configuration for a given \emph{UE Location Pattern}, under a specific PA model. This can be done based on the RL loop, similar to what we proposed in~\cite{HOFFMANN2021232}: UEs report their locations, to be recognized by REM as \emph{UE Location Pattern}, and are considered as \emph{State}, based on which REM decides to select a certain configuration of serving clusters, considered as \emph{Action}, after that EE data is collected and stored in REM constituing a \emph{Reward} associated with the current \emph{NO AP} and \emph{UE Location Pattern}. Then the RL cycle repeats. The procedure is summarized in Fig.~\ref{fig:rl}}

\section{Results}

\begin{figure}[t!]
    \centering
    \begin{subfigure}[t]{0.45\textwidth}
        \centering
        \includegraphics[width=1\linewidth]{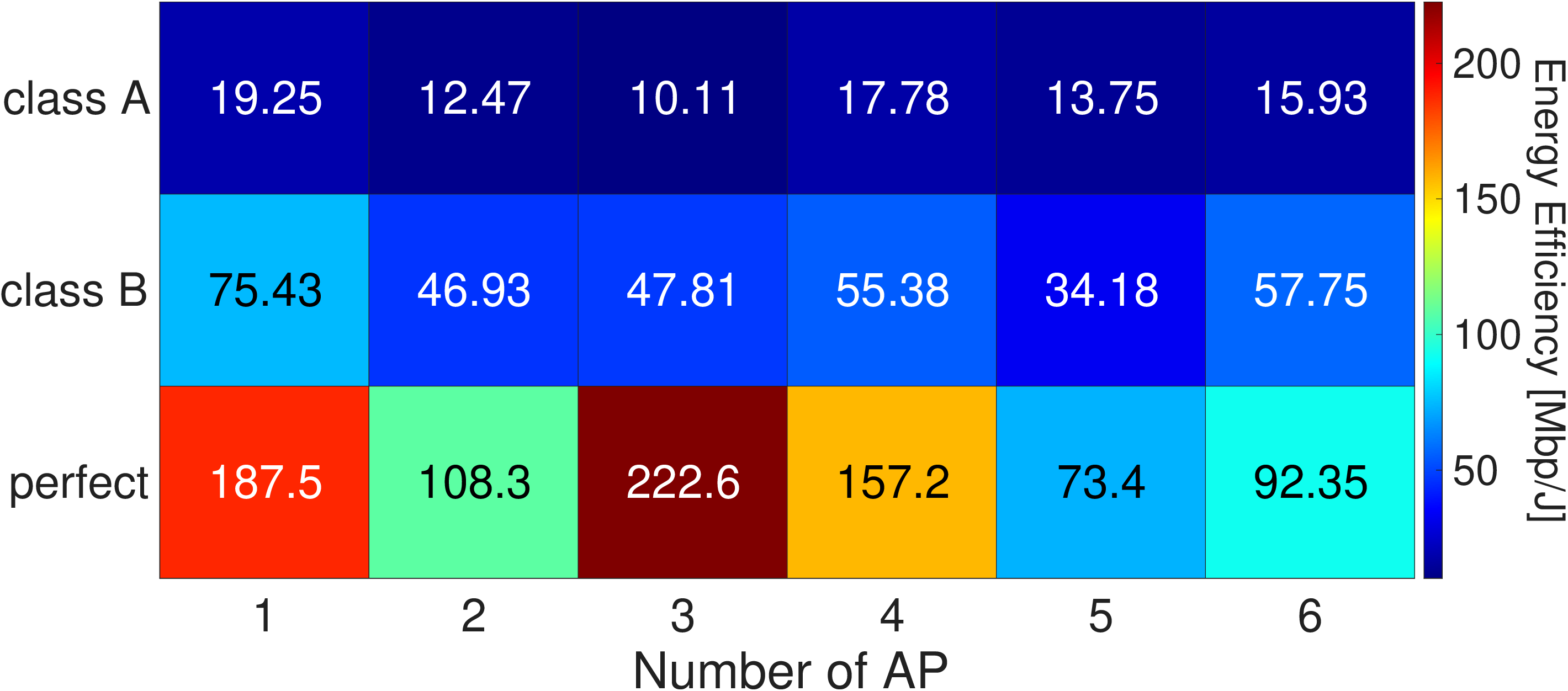}
        \caption{UE location pattern 1.}
        \label{fig:ee_1}
    \end{subfigure}
    ~
    \begin{subfigure}[t]{0.45\textwidth}
        \centering
        \includegraphics[width=1\linewidth]{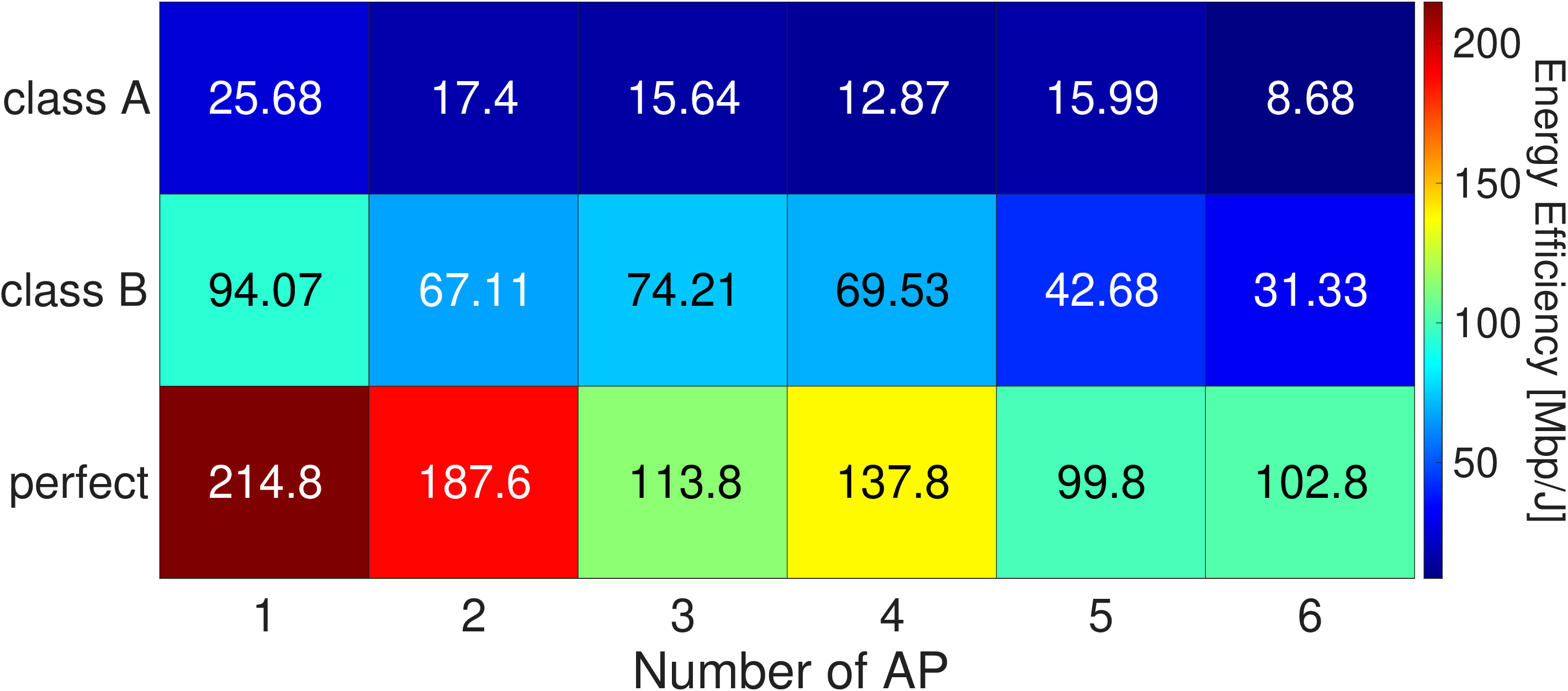}
        \caption{UE location pattern 2.}
        \label{fig:ee_2}
    \end{subfigure}
    \caption{REM entries obtained for 2 exmaple UE location patterns, and 3 models of PA (\emph{class A}, \emph{class B}, and \emph{perfect}).}
\end{figure}

To show how the proposed REM can be created, we use the system-level simulator of the  UCCF M-MIMO network proposed by us in~\cite{hoffmann2024}. It is based on the 3D-RT radio channel model, provides a dedicated space-time-frequency radio resource scheduler, Zero-Forcing precoder, Modulation and Coding Schemas (MCS) allocation, and most importantly, takes into account the nonlinear characteristics of PA by implementing a soft-limiter model. The simulation scenario considers an urban environment with a macro AP equipped with 128 antennas and 46~dBm of maximum transmit power, and 5 micro APs equipped with 32 antennas and 30 dBm of maximum transmit power. In both cases, the maximum transmit power is considered to be the saturation power of the soft-limiter PA installed at APs. To limit the impact of nonlinear distortion, every PA operates at 6~dB Input Back-Off (IBO). The remaining simulation parameters can be found in~\cite{hoffmann2024}. 
We created REM entries by examining the serving cluster configuration for 2 randomly and independently generated \emph{UE Location Patterns} of 40 UEs. To reflect the impact of hardware on EE, the creation of REM entries was repeated, in 3 PA models~\cite{Kryszkiewicz2023}: \emph{class A} (consuming a fixed amount of power), \emph{class B} (consuming an amount of power dependent on the maximum transmit power and IBO), and \emph{perfect} (whose power consumption is equal to the radiated waveform power). 
Based on that, for each PA model, we obtained two example REM entries as depicted in Fig.~\ref{fig:ee_1} and Fig.~\ref{fig:ee_2}, respectively. First observation is that there is a huge about 11-fold difference in EE between using \emph{class A} and \emph{perfect}, PA. Although this is expected, the second interesting observation is that serving a UE with multiple APs can only provide EE improvement under \emph{perfect} PA, for \emph{class A}, and \emph{class B}, it is best to stick to the state-of-the-art approach with serving a UE with the single strongest AP. Even for a \emph{perfect} PA, the configuration of the serving cluster depends on the \emph{UE Location Pattern}, i.e., for \emph{UE Location Pattern 1}, serving UEs with 3 APs is the most EE configuration (almost 19\% improvement over the configuration with 1 AP), while for \emph{UE Location Pattern 2}, a single AP is the best. This justifies the need of the proposed REM. 

\begin{figure}[htbp]
  \centering
  \includegraphics[trim=0cm 0cm 0cm 0cm, clip=true,width=0.45\textwidth]{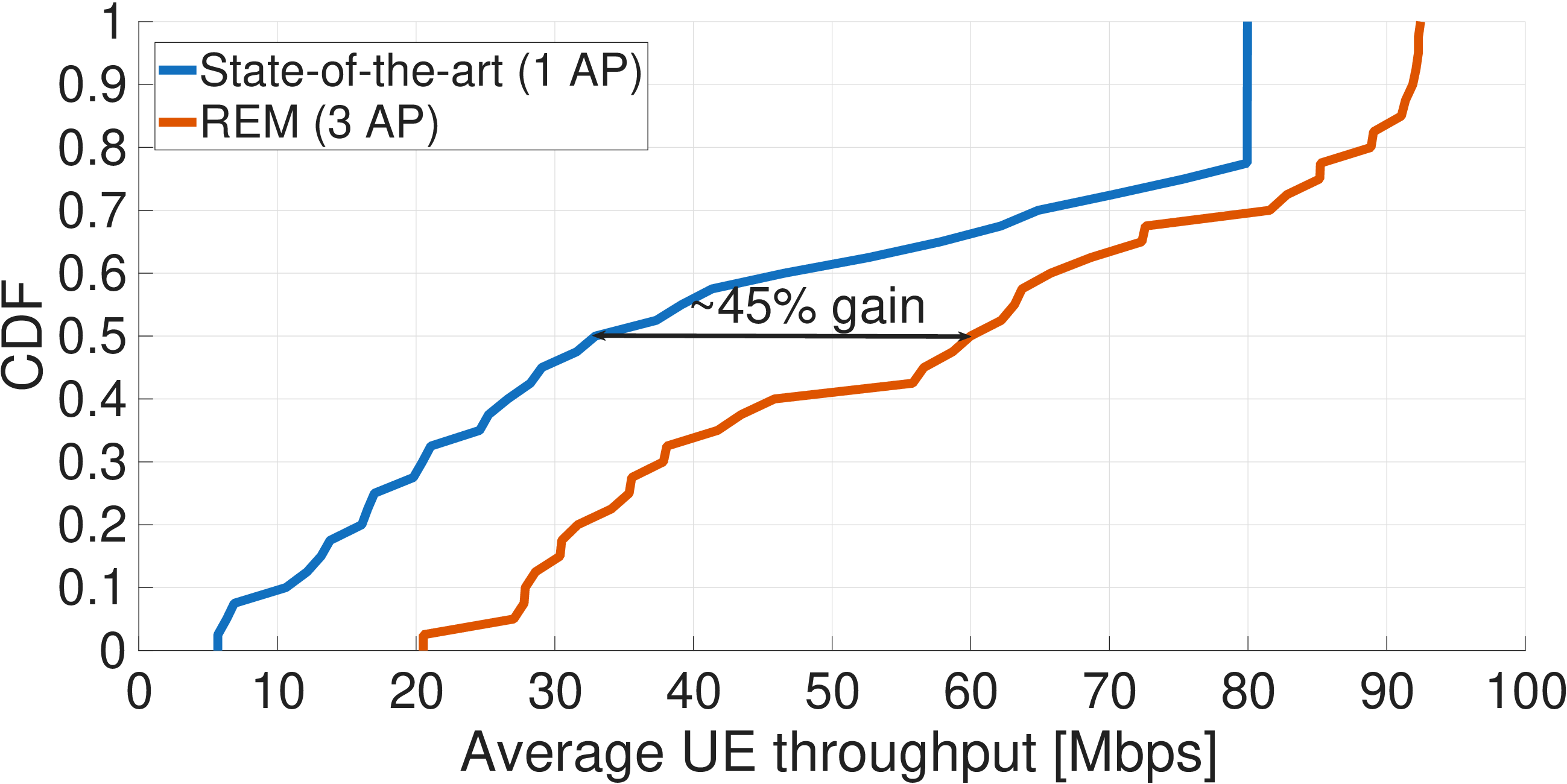}
  \caption{CDF of average UE throughput for \emph{UE Location Pattern 1}, \emph{perfect} PA model and serving cluster configuration of 1 AP (state-of-the-art), and 3 AP (highest EE in REM).}
  \label{fig:cdf}
\end{figure}

{\MH While the EE gain from applying more efficient PA is obvious (denominator of EE definition), additional gain may be achieved from a proper serving cluster by increasing numerator of EE. This is the case for \emph{UE Location Pattern 1} and serving UEs with 3 APs (according to REM-based optimization) instead of the state-of-the-art approach with a single one. The related Cumulative Distribution Function (CDF) of average UE throughput is depicted in Fig.~\ref{fig:cdf}. It can be seen that for this particular set of UE positions utilizaiton of UCCF M-MIMO can improve the throughput of each UE, with a median gain of about 45\%. This is another aspect of REM, using its internal location-dependent data to identify \emph{UE Location Patterns} like this, where extension of serving clusters brings high QoS gains.}

\section{Conclusions and Future Work}

The extension of the UCCF M-MIMO network with REM allows for improving its EE by up to 19\%, with selecting the proper serving cluster configuration for the current \emph{UE Location Pattern} and PAs installed at APs. The natural evolution of the proposed concept is to incorporate AI/ML analysis on top of the REM data.

\section*{Acknowledgements}

Marcin Hoffmann was funded by the Polish National Science Centre, project no. 2022/45/N/ST7/01930. Pawel Kryszkiewicz was funded by the Polish National Science Centre, project no. 2021/41/B/ST7/00136.

\bibliography{references} 
\bibliographystyle{IEEEtran}

\end{document}